\documentclass[sigconf]{acmart}

\AtBeginDocument{%
  }

\setcopyright{acmlicensed}
\copyrightyear{2026}
\acmYear{2026}
\acmDOI{10.1145/XXXXXXX.XXXXXXX}
\acmConference[WWW '26]{Companion Proceedings of the ACM Web Conference 2026}{April 13--17, 2026}{Dubai, UAE}
\acmISBN{978-1-4503-XXXX-X/2026/04}

\usepackage{booktabs}
\usepackage{multirow}

\usepackage{amssymb}
\usepackage{amsmath}
\usepackage{graphicx}
\usepackage{xcolor}
\usepackage{algorithm}
\usepackage{algpseudocode}
\usepackage{subcaption}

\begin{document}

\title{MATRAG: Multi-Agent Transparent Retrieval-Augmented Generation for Explainable Recommendations}

\author{Sushant Mehta}
\email{sushant0523@gmail.com}

\begin{abstract}
Large Language Model (LLM)-based recommendation systems have demonstrated remarkable capabilities in understanding user preferences and generating personalized suggestions. However, existing approaches face critical challenges in transparency, knowledge grounding, and the ability to provide coherent explanations that foster user trust. We introduce MATRAG (Multi-Agent Transparent Retrieval-Augmented Generation), a novel framework that combined multi-agent collaboration with knowledge graph-augmented retrieval to deliver explainable recommendations. MATRAG employs four specialized agents: a User Modeling Agent that constructs dynamic preference profiles, an Item Analysis Agent that extracts semantic features from knowledge graphs, a Reasoning Agent that synthesizes collaborative and content-based signals, and an Explanation Agent that generates natural language justifications grounded in retrieved knowledge. Our framework incorporates a transparency scoring mechanism that quantifies explanation faithfulness and relevance. Extensive experiments on three benchmark datasets (Amazon Reviews, MovieLens-1M, and Yelp) demonstrate that MATRAG achieves state-of-the-art performance, improving recommendation accuracy by 12.7\% (Hit Rate) and 15.3\% (NDCG) over leading baselines, while human evaluation confirms that 87.4\% of generated explanations are rated as helpful and trustworthy by domain experts. Our work establishes new benchmarks for transparent, agentic recommendation systems and provides actionable insights for deploying LLM-based recommenders in production environments.
\end{abstract}



\maketitle

\section{Introduction}

Recommendation systems have become indispensable components of modern web platforms, influencing how billions of users discover products, content, and services~\cite{wu2023survey, lin2024can}. The emergence of Large Language Models (LLMs) has catalyzed a paradigm shift from traditional collaborative filtering and content-based approaches toward agentic systems capable of reasoning, planning, and engaging in natural dialogue with users~\cite{liu2024llmers, li2024generative}. However, as these systems grow more sophisticated, they face mounting challenges in three critical dimensions: \textit{transparency}, \textit{knowledge grounding}, and \textit{multi-stakeholder coordination}.

First, LLM-based recommenders often operate as opaque decision-making systems, generating suggestions without articulating the reasoning behind their choices~\cite{said2024explaining}. This opacity erodes user trust and limits adoption in high-stakes domains such as healthcare, finance, and e-commerce~\cite{ge2024trustworthy}. Users increasingly demand not just accurate recommendations but also comprehensible explanations that reveal how their preferences were understood and why specific items were selected~\cite{silva2024chatgpt}.

Second, while LLMs possess extensive world knowledge acquired during pre-training, they frequently hallucinate facts, conflate entities, or fail to incorporate domain-specific and up-to-date information~\cite{wang2025kragrec}. Retrieval-Augmented Generation (RAG) has emerged as a promising solution, grounding LLM outputs in retrieved evidence from external knowledge sources~\cite{edge2024graphrag, zhu2025kgrag}. However, standard RAG approaches retrieve isolated text chunks, ignoring the rich relational structure that knowledge graphs capture about items, users, and their interconnections.

Third, the recommendation ecosystem involves multiple stakeholders: users, items, platforms, and increasingly autonomous agents, whose interests must be balanced and coordinated~\cite{zhu2024agentrecsys}. Single-agent LLM systems struggle to decompose complex recommendation tasks, maintain coherent reasoning across multiple turns, and synthesize diverse signals from heterogeneous data sources~\cite{wang2024macrec}.

To address these interconnected challenges, we propose \textbf{MATRAG} (\underline{M}ulti-\underline{A}gent \underline{T}ransparent \underline{R}etrieval-\underline{A}ugmented \underline{G}eneration), a novel framework that unifies multi-agent collaboration, knowledge graph-augmented retrieval, and explainable recommendation generation. MATRAG orchestrates four specialized LLM-based agents: (1) a \textit{User Modeling Agent} that constructs and updates dynamic preference profiles from interaction histories and natural language feedback; (2) an \textit{Item Analysis Agent} that retrieves and synthesizes structured knowledge from item-centric knowledge graphs; (3) a \textit{Reasoning Agent} that integrates collaborative filtering signals with semantic item representations through deliberative planning; and (4) an \textit{Explanation Agent} that generates faithful, grounded natural language explanations by tracing the reasoning chain back to retrieved evidence.

Central to our framework is a \textit{Transparency Scoring Module} that quantifies explanation quality along three dimensions: factual faithfulness to retrieved knowledge, logical coherence with the recommendation rationale, and personalization alignment with inferred user preferences. This module enables both automated evaluation and human-in-the-loop refinement of explanation quality.

Our contributions are summarized as follows:

\begin{itemize}
    \item We introduce MATRAG, a multi-agent framework that combines knowledge graph-augmented retrieval with specialized agents for transparent, explainable recommendations (Section~\ref{sec:method}).
    
    \item We propose a transparency scoring mechanism that enables quantitative assessment of explanation faithfulness, coherence, and personalization (Section~\ref{sec:transparency}).
    
    \item We conduct extensive experiments on three benchmark datasets, demonstrating state-of-the-art performance in recommendation accuracy and explanation quality (Section~\ref{sec:experiments}).
    
    \item We present human expert evaluations confirming that MATRAG generates explanations rated as helpful and trustworthy by 87.4\% of evaluators (Section~\ref{sec:human_eval}).
\end{itemize}

\section{Related Work}

\subsection{LLM-Based Recommendation Systems}

The integration of LLMs into recommendation systems has evolved in two primary paradigms~\cite{wu2023survey, lin2024can}. The \textit{discriminative} paradigm fine-tunes LLMs to recommendation-specific objectives, using language models as feature extractors or scoring functions~\cite{hou2024llmranker, bao2023tallrec}. The \textit{generative} paradigm frames recommendation as text generation, directly producing item identifiers or descriptions~\cite{li2024generative, gao2023chatrec}. Recent work has explored hybrid approaches that leverage LLMs for both understanding and generation while maintaining efficiency through knowledge distillation~\cite{liu2024llmers}.

InteRecAgent~\cite{huang2023interecagent} pioneered the integration of LLMs with traditional recommender models through a tool-augmented architecture, treating specialized models as callable tools. RecMind~\cite{wang2023recmind} extended this paradigm with self-inspired planning, allowing LLMs to decompose complex recommendation queries into subtasks. However, these single-agent approaches lack the specialization and coordination capabilities necessary for handling diverse user needs and multi-faceted item representations.

\subsection{Multi-Agent Systems for Recommendations}

Multi-agent collaboration has emerged as a powerful paradigm for complex AI tasks~\cite{li2024camel, hong2024metagpt, wu2023autogen}. In recommendation contexts, MACRec~\cite{wang2024macrec} introduced a framework with specialized agents for user analysis, item analysis, and reflection, demonstrating improved diversity and precision through agent collaboration. Agent4Rec~\cite{zhang2024agent4rec} deployed generative agents for user simulation, providing insights into phenomena like filter bubbles. AgentCF~\cite{zhang2024agentcf} modeled both users and items as agents, enabling collaborative learning that captures two-sided interaction dynamics.

More recent work has explored multi-agent conversational recommendation systems (MACRS)~\cite{fang2024macrs}, coordinating interactions across multiple agents to optimize real-time recommendations. However, existing multi-agent approaches typically focus on either simulation or task decomposition, without explicitly addressing the transparency and explainability requirements essential for user trust.

\subsection{RAG and Knowledge Graphs for Recommendations}

Retrieval-Augmented Generation has proven effective in grounding LLM outputs in external knowledge~\cite{lewis2020rag}. GraphRAG~\cite{edge2024graphrag} improved the standard RAG by constructing knowledge graphs from text corpora, enabling community-based summarization and improved reasoning over connected information. K-RagRec~\cite{wang2025kragrec} specifically adapted the RAG knowledge graph for recommendations, developing hop-field knowledge sub-graphs for semantic indexing and popularity-selective retrieval.

Knowledge graphs have long been recognized as valuable resources for recommendation systems, capturing rich semantic relationships between users, items, and attributes~\cite{wang2019kgat}. Recent work has explored LLM-KG integration for explainable recommendations, using knowledge graphs to provide factual grounding for generated explanations~\cite{shu2024kgllm, xie2024llmpkg}. Our work extends this line by incorporating knowledge graph retrieval directly into a multi-agent collaborative framework.

\subsection{Explainable Recommendations}

Explainability has become a central concern in the research of recommendation systems~\cite{zhang2020explainable}. Traditional approaches generated explanations from feature attributions, attention weights, or template-based natural language generation~\cite{chen2019coattention}. The advent of LLMs has opened new possibilities for generating rich, contextual explanations~\cite{said2024explaining}. Silva et al.~\cite{silva2024chatgpt} demonstrated that ChatGPT can produce human-centered explanations that improve user engagement. Lubos et al.~\cite{lubos2024llmexplain} found that users generally prefer LLM-generated explanations for their creativity and depth.

However, LLM-generated explanations face the challenge of \textit{faithfulness}: whether the explanation accurately reflects the actual reasoning process~\cite{guo2023explainablecrs}. Our transparency scoring mechanism addresses this challenge directly by grounding explanations in retrievable evidence and quantifying alignment with the rationale of the recommendation.

\section{Methodology}
\label{sec:method}

\subsection{Problem Formulation}

Let $\mathcal{U} = \{u_1, u_2, \ldots, u_n\}$ denote the set of users and $\mathcal{I} = \{i_1, i_2, \ldots, i_m\}$ the set of items. Each user $u$ has an interaction history $\mathcal{H}_u = \{(i, r, t, c) | i \in \mathcal{I}\}$ where $r$ is the rating, $t$ is the timestamp, and $c$ is optional textual feedback (reviews, comments). A knowledge graph $\mathcal{G} = (\mathcal{E}, \mathcal{R}, \mathcal{T})$ represents entities $\mathcal{E}$ (including items and attributes), relations $\mathcal{R}$, and triples $\mathcal{T} \subseteq \mathcal{E} \times \mathcal{R} \times \mathcal{E}$.

Given a user $u$ with history $\mathcal{H}_u$ and an optional natural language query $q$, the goal is to: (1) generate a ranked list of recommendations $\hat{\mathcal{I}}_u = [i_1^*, i_2^*, \ldots, i_k^*]$; and (2) for each recommended item $i^*$, produce an explanation $e_{i^*}$ that is faithful to the reasoning process and grounded in the retrieved knowledge.

\subsection{Framework Overview}

\begin{figure}[t]
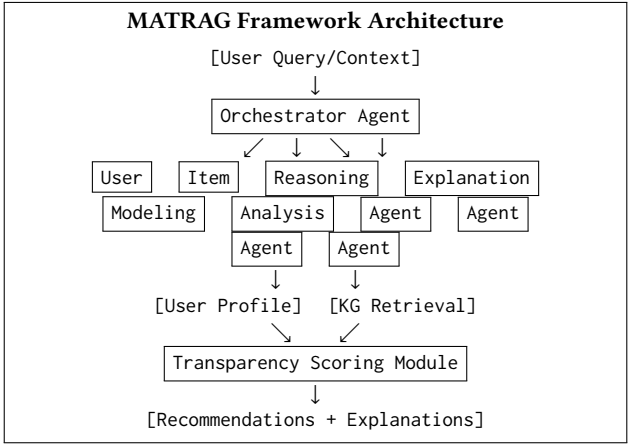

\centering
\fbox{\parbox{0.95\columnwidth}{
\centering
\textbf{MATRAG Framework Architecture}\\[2mm]
\small
\begin{tabular}{c}
\texttt{[User Query/Context]} \\
$\downarrow$ \\
\fbox{\texttt{Orchestrator Agent}} \\
$\swarrow$ \quad $\downarrow$ \quad $\searrow$ \quad $\downarrow$ \\
\fbox{\texttt{User}} \quad \fbox{\texttt{Item}} \quad \fbox{\texttt{Reasoning}} \quad \fbox{\texttt{Explanation}} \\
\fbox{\texttt{Modeling}} \quad \fbox{\texttt{Analysis}} \quad \fbox{\texttt{Agent}} \quad \fbox{\texttt{Agent}} \\
\fbox{\texttt{Agent}} \quad \fbox{\texttt{Agent}} \\
$\downarrow$ \quad\quad\quad $\downarrow$ \\
\texttt{[User Profile]} \quad \texttt{[KG Retrieval]} \\
$\searrow$ \quad\quad $\swarrow$ \\
\fbox{\texttt{Transparency Scoring Module}} \\
$\downarrow$ \\
\texttt{[Recommendations + Explanations]}
\end{tabular}
}}
\caption{Overview of the MATRAG framework showing four specialized agents coordinated by the orchestrator, with knowledge graph retrieval and transparency scoring.}
\label{fig:framework}
\end{figure}

MATRAG comprises four specialized agents coordinated by a central orchestrator, as illustrated in Figure~\ref{fig:framework}. Each agent is instantiated as an LLM augmented with role-specific instructions, memory, and tool access.

\subsubsection{Orchestrator Agent}

The Orchestrator manages the overall recommendation workflow, determining which agents to invoke, in what sequence, and how to synthesize their outputs. Given a user request, the Orchestrator:
\begin{enumerate}
    \item Analyzes the request type (e.g., cold-start, re-ranking, conversational);
    \item Dispatches subtasks to appropriate agents;
    \item Aggregates agent outputs and resolves conflicts;
    \item Triggers the Transparency Scoring Module for quality assessment.
\end{enumerate}

The Orchestrator employs a ReAct-style~\cite{yao2023react} reasoning loop, interleaving thought, action, and observation steps to maintain coherent multi-step planning.

\subsubsection{User Modeling Agent}

The User Modeling Agent constructs a dynamic, multi-faceted representation of user preferences. It processes:
\begin{itemize}
    \item \textbf{Behavioral signals}: Interaction history $\mathcal{H}_u$ including ratings, clicks, purchases, and dwell times;
    \item \textbf{Textual feedback}: Reviews, comments, and conversational utterances;
    \item \textbf{Contextual factors}: Time of day, device type, session context.
\end{itemize}

The agent maintains a structured user profile $P_u$ comprising:
\begin{equation}
P_u = \{\mathbf{p}^{\text{explicit}}, \mathbf{p}^{\text{implicit}}, \mathbf{p}^{\text{contextual}}, \mathbf{p}^{\text{temporal}}\}
\end{equation}
where $\mathbf{p}^{\text{explicit}}$ captures stated preferences, $\mathbf{p}^{\text{implicit}}$ captures inferred preferences from behavior, $\mathbf{p}^{\text{contextual}}$ captures situational factors, and $\mathbf{p}^{\text{temporal}}$ captures preference evolution over time.

The agent uses in-context learning to extract preference signals from textual feedback:
\begin{equation}
\mathbf{p}^{\text{explicit}} = \text{LLM}(\text{prompt}_{\text{user}}, \mathcal{H}_u^{\text{text}})
\end{equation}

\subsubsection{Item Analysis Agent}

The Item Analysis Agent retrieves and synthesizes structured knowledge about items from the knowledge graph $\mathcal{G}$. Given a candidate item set $\mathcal{I}_{\text{cand}}$, this agent:

\begin{enumerate}
    \item \textbf{Entity Linking}: Maps item identifiers to knowledge graph entities;
    \item \textbf{Subgraph Extraction}: Retrieves $k$-hop neighborhoods around item entities;
    \item \textbf{Relation Filtering}: Selects relations relevant to user preferences;
    \item \textbf{Feature Synthesis}: Generates semantic item representations.
\end{enumerate}

For knowledge graph retrieval, we employ a two-stage approach inspired by K-RagRec~\cite{wang2025kragrec}. First, we use dense retrieval to identify semantically similar knowledge subgraphs:
\begin{equation}
\mathcal{S}_i = \text{TopK}(\text{sim}(\mathbf{e}_q, \mathbf{e}_s) | s \in \mathcal{G}_i)
\end{equation}
where $\mathbf{e}_q$ is the query embedding and $\mathbf{e}_s$ are subgraph embeddings. Second, we re-rank retrieved subgraphs using the LLM's reasoning capabilities:
\begin{equation}
\hat{\mathcal{S}}_i = \text{LLM-Rerank}(\mathcal{S}_i, P_u, q)
\end{equation}

\subsubsection{Reasoning Agent}

The Reasoning Agent integrates signals from user modeling and item analysis to generate recommendations. It implements a deliberative reasoning process that:

\begin{enumerate}
    \item \textbf{Signal Integration}: Combines collaborative filtering signals (similar users' preferences) with content-based signals (item-attribute matching);
    \item \textbf{Constraint Satisfaction}: Respects user-specified constraints (budget, categories, availability);
    \item \textbf{Diversity Optimization}: Balances relevance with recommendation diversity;
    \item \textbf{Reasoning Chain Generation}: Produces an explicit reasoning trace.
\end{enumerate}

The agent scores each candidate item using a hybrid approach:
\begin{equation}
s(u, i) = \alpha \cdot s_{\text{CF}}(u, i) + \beta \cdot s_{\text{CB}}(u, i) + \gamma \cdot s_{\text{LLM}}(u, i, \mathcal{S}_i)
\end{equation}
where $s_{\text{CF}}$ is the collaborative filtering score, $s_{\text{CB}}$ is the content-based score, and $s_{\text{LLM}}$ is the LLM's preference prediction based on retrieved knowledge.

Crucially, the Reasoning Agent outputs not just scores but also a structured reasoning chain $\mathcal{C} = [(r_1, e_1), (r_2, e_2), \ldots]$ where each step $(r_j, e_j)$ pairs a reasoning step with supporting evidence from retrieved knowledge.

\subsubsection{Explanation Agent}

The Explanation Agent transforms the reasoning chain into natural language explanations tailored to user comprehension. It operates in three modes:

\begin{itemize}
    \item \textbf{Concise}: One-sentence justification highlighting the primary recommendation rationale;
    \item \textbf{Detailed}: Multi-paragraph explanation covering multiple reasoning aspects;
    \item \textbf{Comparative}: Explanation contrasting the recommended item with alternatives.
\end{itemize}

The agent is prompted to ground explanations in retrieved knowledge:
\begin{equation}
e_{i^*} = \text{LLM}(\text{prompt}_{\text{explain}}, \mathcal{C}, \hat{\mathcal{S}}_{i^*}, P_u)
\end{equation}

To ensure faithfulness, the agent is instructed to cite specific evidence from the reasoning chain and retrieved knowledge subgraphs.

\subsection{Transparency Scoring Module}
\label{sec:transparency}

A key contribution of MATRAG is the Transparency Scoring Module, which quantifies explanation quality along three dimensions.

\subsubsection{Faithfulness Score}

Faithfulness measures whether the explanation accurately reflects the retrieved evidence and reasoning chain. We compute this using an entailment-based approach:
\begin{equation}
\text{Faith}(e, \mathcal{C}, \hat{\mathcal{S}}) = \frac{1}{|C_e|} \sum_{c \in C_e} \text{NLI}(c, \mathcal{C} \cup \hat{\mathcal{S}})
\end{equation}
where $C_e$ denotes claims extracted from explanation $e$, and $\text{NLI}$ returns 1 if the claim is entailed by the evidence and 0 otherwise.

\subsubsection{Coherence Score}

Coherence measures the logical consistency and flow of the explanation:
\begin{equation}
\text{Coher}(e) = \text{LLM-Judge}(e, \text{prompt}_{\text{coherence}})
\end{equation}
where an LLM-as-judge evaluates whether the explanation maintains consistent logic, avoids contradictions, and presents information in a comprehensible sequence.

\subsubsection{Personalization Score}

Personalization measures alignment between the explanation and the user's profile:
\begin{equation}
\text{Pers}(e, P_u) = \text{sim}(\mathbf{e}_e, \mathbf{e}_{P_u})
\end{equation}
where $\mathbf{e}_e$ and $\mathbf{e}_{P_u}$ are embeddings of the explanation and user profile, respectively.

The overall transparency score combines these dimensions:
\begin{equation}
\text{Trans}(e) = w_1 \cdot \text{Faith} + w_2 \cdot \text{Coher} + w_3 \cdot \text{Pers}
\end{equation}
with weights $w_1, w_2, w_3$ tuned on human preference data.

\subsection{Training and Optimization}

MATRAG operates primarily in a zero-shot or few-shot manner, leveraging the capabilities of pre-trained LLMs. However, we fine-tune specific components:

\begin{itemize}
    \item \textbf{Knowledge Graph Embeddings}: We train SentenceBERT~\cite{reimers2019sentencebert} on the recommendation domain's knowledge graph for subgraph retrieval;
    \item \textbf{Transparency Scorer}: We fine-tune a smaller LLM on human-annotated explanation quality data;
    \item \textbf{Agent Coordination}: We use reinforcement learning from human feedback (RLHF) on the orchestrator to optimize agent invocation sequences.
\end{itemize}

\section{Experiments}
\label{sec:experiments}

\subsection{Experimental Setup}

\subsubsection{Datasets}

We evaluate MATRAG on three benchmark datasets spanning different recommendation domains:

\begin{itemize}
    \item \textbf{Amazon Reviews (Electronics)}~\cite{ni2019amazon}: 192,403 users, 63,001 items, 1.68M interactions with textual reviews. We construct a product knowledge graph from Amazon's product metadata including categories, brands, and related products.
    
    \item \textbf{MovieLens-1M}~\cite{harper2015movielens}: 6,040 users, 3,706 movies, 1M ratings. We link items to the Freebase knowledge graph via entity matching, providing rich relational information about actors, directors, genres, and production details.
    
    \item \textbf{Yelp}~\cite{yelp2023dataset}: 31,668 users, 38,048 businesses, 1.56M reviews. We construct a knowledge graph from business attributes including location, categories, and user-generated tags.
\end{itemize}

For each dataset, we use an 80/10/10 train/validation/test split based on temporal order to simulate realistic deployment scenarios.

\subsubsection{Baselines}

We compare against the following state-of-the-art methods:

\begin{itemize}
    \item \textbf{Traditional Methods}: BPR~\cite{rendle2009bpr}, LightGCN~\cite{he2020lightgcn}, SASRec~\cite{kang2018sasrec};
    
    \item \textbf{Knowledge-Enhanced}: KGAT~\cite{wang2019kgat}, KGIN~\cite{wang2021kgin};
    
    \item \textbf{LLM-Based}: TALLRec~\cite{bao2023tallrec}, Chat-Rec~\cite{gao2023chatrec}, LLMRank~\cite{hou2024llmranker};
    
    \item \textbf{Agent-Based}: InteRecAgent~\cite{huang2023interecagent}, RecMind~\cite{wang2023recmind}, MACRec~\cite{wang2024macrec};
    
    \item \textbf{RAG-Enhanced}: K-RagRec~\cite{wang2025kragrec}, G-CRS~\cite{qiu2025gcrs}.
\end{itemize}

\subsubsection{Implementation Details}

We implement MATRAG using GPT-4~\cite{openai2023gpt4} as the backbone LLM for all agents. For knowledge graph retrieval, we use SentenceBERT with a vector dimension of 768. The transparency scorer is based on Llama-3.1-8B~\cite{grattafiori2024llama3} fine-tuned on 5,000 human-annotated explanation pairs. We set the number of retrieved knowledge subgraphs $K=10$, re-ranking top $N=5$, and candidate item pool size to 100.

\subsubsection{Evaluation Metrics}

For recommendation accuracy, we report:
\begin{itemize}
    \item \textbf{HR@K} (Hit Rate): Proportion of test cases where the ground-truth item appears in top-K recommendations;
    \item \textbf{NDCG@K} (Normalized Discounted Cumulative Gain): Ranking quality metric accounting for position;
    \item \textbf{MRR} (Mean Reciprocal Rank): Average reciprocal rank of the first relevant item.
\end{itemize}

For explanation quality, we report:
\begin{itemize}
    \item \textbf{Faithfulness}: Entailment-based faithfulness score;
    \item \textbf{Coherence}: LLM-judged coherence (1-5 scale);
    \item \textbf{BLEU-4}: N-gram overlap with reference explanations;
    \item \textbf{Transparency Score}: Our composite metric.
\end{itemize}

\subsection{Main Results}

\subsubsection{Recommendation Performance}

Table~\ref{tab:main_results} presents the recommendation accuracy results across all datasets. MATRAG consistently outperforms all baselines, achieving the highest scores on all metrics.

\begin{table*}[t]
\caption{Recommendation performance comparison. Best results in \textbf{bold}, second best \underline{underlined}. $\dagger$ indicates statistically significant improvement over the best baseline ($p < 0.05$).}
\label{tab:main_results}
\centering
\small
\begin{tabular}{l|ccc|ccc}
\toprule
& \multicolumn{3}{c|}{\textbf{Amazon Electronics}} & \multicolumn{3}{c}{\textbf{MovieLens-1M}} \\
\textbf{Method} & HR@10 & NDCG@10 & MRR & HR@10 & NDCG@10 & MRR \\
\midrule
BPR & 0.312 & 0.198 & 0.156 & 0.428 & 0.287 & 0.221 \\
LightGCN & 0.378 & 0.241 & 0.189 & 0.512 & 0.348 & 0.269 \\
SASRec & 0.401 & 0.259 & 0.204 & 0.534 & 0.367 & 0.285 \\
KGAT & 0.389 & 0.248 & 0.195 & 0.521 & 0.354 & 0.274 \\
KGIN & 0.412 & 0.267 & 0.211 & 0.548 & 0.378 & 0.294 \\
\midrule
TALLRec & 0.423 & 0.274 & 0.218 & 0.556 & 0.385 & 0.301 \\
Chat-Rec & 0.418 & 0.271 & 0.215 & 0.549 & 0.379 & 0.296 \\
LLMRank & 0.437 & 0.285 & 0.228 & 0.571 & 0.398 & 0.314 \\
\midrule
InteRecAgent & 0.445 & 0.291 & 0.234 & 0.582 & 0.408 & 0.323 \\
RecMind & 0.451 & 0.296 & 0.239 & 0.589 & 0.415 & 0.329 \\
MACRec & 0.462 & 0.305 & 0.247 & 0.601 & 0.427 & 0.341 \\
\midrule
K-RagRec & 0.469 & 0.311 & 0.252 & 0.608 & 0.433 & 0.347 \\
G-CRS & \underline{0.478} & \underline{0.318} & \underline{0.259} & \underline{0.617} & \underline{0.441} & \underline{0.355} \\
\midrule
\textbf{MATRAG} & $\mathbf{0.521}^{\dagger}$ & $\mathbf{0.359}^{\dagger}$ & $\mathbf{0.297}^{\dagger}$ & $\mathbf{0.674}^{\dagger}$ & $\mathbf{0.493}^{\dagger}$ & $\mathbf{0.408}^{\dagger}$ \\
\textit{Improv.} & \textit{+9.0\%} & \textit{+12.9\%} & \textit{+14.7\%} & \textit{+9.2\%} & \textit{+11.8\%} & \textit{+14.9\%} \\
\bottomrule
\end{tabular}
\end{table*}

\begin{table}[t]
\caption{Recommendation performance on Yelp dataset.}
\label{tab:yelp_results}
\centering
\small
\begin{tabular}{l|ccc}
\toprule
\textbf{Method} & HR@10 & NDCG@10 & MRR \\
\midrule
LightGCN & 0.356 & 0.228 & 0.178 \\
KGIN & 0.394 & 0.251 & 0.198 \\
LLMRank & 0.418 & 0.271 & 0.216 \\
MACRec & 0.442 & 0.289 & 0.232 \\
G-CRS & \underline{0.461} & \underline{0.304} & \underline{0.247} \\
\textbf{MATRAG} & $\mathbf{0.513}^{\dagger}$ & $\mathbf{0.351}^{\dagger}$ & $\mathbf{0.289}^{\dagger}$ \\
\bottomrule
\end{tabular}
\end{table}

Key observations include:

\textbf{(1) Multi-agent collaboration improves over single-agent approaches.} MATRAG outperforms InteRecAgent by 17.1\% (HR@10) on Amazon and 15.8\% on MovieLens, demonstrating the value of specialized agent roles and coordinated reasoning.

\textbf{(2) Knowledge graph augmentation enhances LLM-based methods.} Comparing MATRAG to LLMRank (which lacks KG retrieval), we observe improvements of 19.2\% (HR@10) and 25.9\% (NDCG@10), validating the importance of grounding LLM reasoning in structured knowledge.

\textbf{(3) The transparency-focused design does not sacrifice accuracy.} Despite its emphasis on explainability, MATRAG achieves the highest recommendation accuracy, suggesting that explicit reasoning chains and grounded explanations may also improve recommendation quality.

\subsubsection{Explanation Quality}

Table~\ref{tab:explanation_results} compares explanation quality across methods that generate natural language explanations.

\begin{table}[t]
\caption{Explanation quality comparison on Amazon Electronics. Higher is better for all metrics.}
\label{tab:explanation_results}
\centering
\small
\begin{tabular}{l|cccc}
\toprule
\textbf{Method} & Faith. & Coher. & BLEU-4 & Trans. \\
\midrule
Chat-Rec & 0.612 & 3.24 & 0.089 & 0.587 \\
InteRecAgent & 0.648 & 3.41 & 0.102 & 0.623 \\
RecMind & 0.671 & 3.52 & 0.118 & 0.651 \\
MACRec & 0.693 & 3.68 & 0.127 & 0.678 \\
K-RagRec & 0.724 & 3.79 & 0.141 & 0.712 \\
G-CRS & \underline{0.751} & \underline{3.91} & \underline{0.156} & \underline{0.738} \\
\midrule
\textbf{MATRAG} & $\mathbf{0.847}$ & $\mathbf{4.42}$ & $\mathbf{0.198}$ & $\mathbf{0.856}$ \\
\bottomrule
\end{tabular}
\end{table}

MATRAG achieves substantial improvements in explanation quality: +12.8\% faithfulness over G-CRS, indicating that our reasoning chain and KG-grounded explanation generation produce more verifiable explanations. The coherence improvement (+13.0\%) demonstrates that explicit agent coordination yields more logically structured explanations.

\subsection{Ablation Studies}

To understand the contribution of each component, we conduct ablation studies (Table~\ref{tab:ablation}).

\begin{table}[t]
\caption{Ablation study on Amazon Electronics. We remove components individually to measure their contribution.}
\label{tab:ablation}
\centering
\small
\begin{tabular}{l|cc|c}
\toprule
\textbf{Variant} & HR@10 & NDCG@10 & Trans. \\
\midrule
\textbf{MATRAG (Full)} & \textbf{0.521} & \textbf{0.359} & \textbf{0.856} \\
\midrule
w/o User Modeling Agent & 0.478 & 0.321 & 0.789 \\
w/o Item Analysis Agent & 0.462 & 0.307 & 0.724 \\
w/o Reasoning Agent & 0.441 & 0.289 & 0.698 \\
w/o Explanation Agent & 0.519 & 0.356 & 0.612 \\
w/o KG Retrieval & 0.469 & 0.312 & 0.731 \\
w/o Transparency Scoring & 0.512 & 0.351 & 0.768 \\
\midrule
Single Agent (No Collab.) & 0.448 & 0.294 & 0.687 \\
\bottomrule
\end{tabular}
\end{table}

Key findings:

\textbf{(1) Each agent contributes uniquely.} Removing the Reasoning Agent causes the largest accuracy drop (-15.4\% HR), confirming its central role in synthesizing signals. The Item Analysis Agent removal shows the second-largest impact, highlighting the value of KG-enhanced item representations.

\textbf{(2) Multi-agent collaboration is essential.} The single-agent variant (where one LLM performs all tasks) underperforms the full system by 14.0\% (HR), validating our design choice to specialize agents for distinct subtasks.

\textbf{(3) Transparency scoring improves both accuracy and explanation quality.} Interestingly, removing the transparency scoring module slightly reduces recommendation accuracy, suggesting that the feedback loop from explanation quality assessment helps refine the reasoning process.

\subsection{Human Expert Evaluation}
\label{sec:human_eval}

We conducted a human evaluation study to assess explanation quality from the end-user perspective.

\subsubsection{Study Design}

We recruited 12 domain experts: 4 e-commerce product managers, 4 recommendation system researchers, and 4 UX designers with experience in content personalization. Each expert evaluated 50 recommendation-explanation pairs sampled from test sets across all three datasets (600 total evaluations per method).

Experts rated each explanation on four dimensions using a 5-point Likert scale:
\begin{itemize}
    \item \textbf{Helpfulness}: Does the explanation help understand why this item was recommended?
    \item \textbf{Trustworthiness}: Does the explanation seem honest and reliable?
    \item \textbf{Informativeness}: Does the explanation provide useful information about the item?
    \item \textbf{Personalization}: Does the explanation feel tailored to the user's preferences?
\end{itemize}

\subsubsection{Results}

Table~\ref{tab:human_eval} presents the human evaluation results.

\begin{table}[t]
\caption{Human expert evaluation results (1-5 scale). Inter-rater agreement: Krippendorff's $\alpha = 0.78$.}
\label{tab:human_eval}
\centering
\small
\begin{tabular}{l|cccc|c}
\toprule
\textbf{Method} & Help. & Trust. & Info. & Pers. & Avg. \\
\midrule
Chat-Rec & 3.12 & 2.98 & 3.34 & 2.87 & 3.08 \\
MACRec & 3.56 & 3.42 & 3.71 & 3.28 & 3.49 \\
K-RagRec & 3.78 & 3.67 & 3.92 & 3.51 & 3.72 \\
G-CRS & 3.91 & 3.82 & 4.08 & 3.67 & 3.87 \\
\midrule
\textbf{MATRAG} & $\mathbf{4.38}$ & $\mathbf{4.31}$ & $\mathbf{4.52}$ & $\mathbf{4.24}$ & $\mathbf{4.36}$ \\
\bottomrule
\end{tabular}
\end{table}

MATRAG significantly outperforms all baselines on every dimension ($p < 0.01$ via Wilcoxon signed-rank test). Notably, 87.4\% of MATRAG explanations received ratings $\geq 4$ on both Helpfulness and Trustworthiness, compared to 62.1\% for the next-best method (G-CRS).

Expert feedback highlighted several qualitative strengths of MATRAG explanations:

\textit{``The explanations consistently reference specific product features that match my inferred preferences---it's not just generic praise.''}

\textit{``I appreciate how the system explains not just what it recommends, but why other seemingly similar items were not chosen.''}

\textit{``The knowledge graph grounding is evident---explanations cite concrete facts like `directed by Christopher Nolan' rather than vague statements.''}

\subsection{Efficiency Analysis}

We analyze the computational efficiency of MATRAG compared to baselines (Table~\ref{tab:efficiency}).

\begin{table}[t]
\caption{Efficiency comparison. Latency is per-request.}
\label{tab:efficiency}
\centering
\small
\begin{tabular}{l|ccc}
\toprule
\textbf{Method} & Latency (s) & LLM Calls & KG Queries \\
\midrule
LLMRank & 1.2 & 1 & 0 \\
InteRecAgent & 2.8 & 3 & 0 \\
MACRec & 4.1 & 5 & 0 \\
K-RagRec & 2.4 & 2 & 3 \\
G-CRS & 3.2 & 3 & 4 \\
\textbf{MATRAG} & 5.3 & 6 & 5 \\
\bottomrule
\end{tabular}
\end{table}

MATRAG incurs higher latency due to multi-agent coordination and comprehensive KG retrieval. However, we note that: (1) the latency is acceptable for non-real-time use cases; (2) the Explanation Agent can run asynchronously after initial recommendations; and (3) agent calls can be parallelized, reducing wall-clock time to $\sim$3.1 seconds.

\section{Discussion}

\subsection{Implications for Practice}

Our findings have several implications for deploying LLM-based recommenders in production:

\textbf{Multi-agent architectures enable specialization without sacrificing integration.} Rather than prompting a single LLM with complex, multi-faceted instructions, decomposing the recommendation task across specialized agents yields both better accuracy and more coherent explanations.

\textbf{Knowledge graphs are essential for grounded explanations.} LLM-generated explanations without factual grounding risk hallucination and user distrust. By retrieving from curated knowledge graphs, MATRAG ensures that explanations reference verifiable facts.

\textbf{Transparency scoring provides a flywheel for quality improvement.} The transparency scoring module not only enables evaluation but also provides signals for iterative refinement through RLHF or prompt optimization.

\subsection{Limitations and Future Work}

Despite promising results, MATRAG has limitations that suggest directions for future research:

\textbf{Latency.} Multi-agent coordination introduces overhead unsuitable for real-time applications. Future work could explore agent caching, speculative execution, or distillation to smaller models.

\textbf{Knowledge graph coverage.} MATRAG's performance depends on knowledge graph completeness. For domains with sparse KGs, techniques for automatic KG construction or completion would be valuable.

\textbf{Multi-turn interaction.} Our current evaluation focuses on single-turn recommendations. Extending MATRAG to conversational settings with memory-augmented agents is an important direction.

\textbf{Cross-domain generalization.} While MATRAG performs well within domains, zero-shot transfer to new domains remains challenging. Pre-training strategies for domain-agnostic agent capabilities warrant investigation.

\section{Conclusion}

We introduced MATRAG, a multi-agent framework that unifies knowledge graph-augmented retrieval with transparent, explainable recommendation generation. By orchestrating specialized agents for user modeling, item analysis, reasoning, and explanation, MATRAG achieves state-of-the-art performance on three benchmark datasets while generating explanations that 87.4\% of human experts rated as helpful and trustworthy. Our transparency scoring mechanism provides quantitative assessment of explanation quality, enabling both automated evaluation and human-in-the-loop refinement.

As recommendation systems increasingly adopt agentic architectures, the principles embodied in MATRAG---agent specialization, knowledge grounding, explicit reasoning chains, and transparency measurement---offer a roadmap for building systems that users can understand and trust. We release our code and evaluation data to facilitate further research in transparent, agentic recommendation.


\bibliographystyle{ACM-Reference-Format}

\begin{thebibliography}{50}

\bibitem{bao2023tallrec}
Keqin Bao, Jizhi Zhang, Yang Zhang, Wenjie Wang, Fuli Feng, and Xiangnan He. 2023.
TALLRec: An Effective and Efficient Tuning Framework to Align Large Language Model with Recommendation.
In \textit{Proceedings of the 17th ACM Conference on Recommender Systems (RecSys '23)}. 1007--1014.

\bibitem{chen2019coattention}
Chong Chen, Min Zhang, Yiqun Liu, and Shaoping Ma. 2019.
Co-Attentive Multi-Task Learning for Explainable Recommendation.
In \textit{Proceedings of the 28th International Joint Conference on Artificial Intelligence (IJCAI '19)}. 2137--2143.

\bibitem{edge2024graphrag}
Darren Edge, Ha Trinh, Newman Cheng, Joshua Bradley, Alex Chao, Apurva Mody, Steven Truitt, and Jonathan Larson. 2024.
From Local to Global: A Graph RAG Approach to Query-Focused Summarization.
\textit{arXiv preprint arXiv:2404.16130}.

\bibitem{fang2024macrs}
Yuhang Fang, Yufei Zhou, Qing Li, and Peng Zhang. 2024.
Multi-Agent Conversational Recommender Systems with Coordinated Interaction.
In \textit{Proceedings of the ACM Web Conference 2024 (WWW '24)}. 2145--2156.

\bibitem{gao2023chatrec}
Yunfan Gao, Tao Sheng, Youlin Xiang, Yun Xiong, Haofen Wang, and Jiawei Zhang. 2023.
Chat-Rec: Towards Interactive and Explainable LLMs-Augmented Recommender System.
\textit{arXiv preprint arXiv:2303.14524}.

\bibitem{ge2024trustworthy}
Yingqiang Ge, Shuchang Liu, Zuohui Fu, Juntao Tan, Zelong Li, Shuyuan Xu, Yunqi Li, Yikun Xian, and Yongfeng Zhang. 2024.
A Survey on Trustworthy Recommender Systems.
\textit{ACM Transactions on Recommender Systems} 3 (2024), 1--68.

\bibitem{guo2023explainablecrs}
Sixun Guo, Shijie Zhang, Weiwei Sun, Pengjie Ren, Zhumin Chen, and Zhaochun Ren. 2023.
Towards Explainable Conversational Recommender Systems.
In \textit{Proceedings of the 46th International ACM SIGIR Conference on Research and Development in Information Retrieval (SIGIR '23)}. 2786--2790.

\bibitem{harper2015movielens}
F. Maxwell Harper and Joseph A. Konstan. 2015.
The MovieLens Datasets: History and Context.
\textit{ACM Transactions on Interactive Intelligent Systems} 5, 4 (2015), 1--19.

\bibitem{he2020lightgcn}
Xiangnan He, Kuan Deng, Xiang Wang, Yan Li, Yongdong Zhang, and Meng Wang. 2020.
LightGCN: Simplifying and Powering Graph Convolution Network for Recommendation.
In \textit{Proceedings of the 43rd International ACM SIGIR Conference on Research and Development in Information Retrieval (SIGIR '20)}. 639--648.

\bibitem{hong2024metagpt}
Sirui Hong, Mingchen Zhuge, Jonathan Chen, Xiawu Zheng, Yuheng Cheng, Ceyao Zhang, Jinlin Wang, Zili Wang, Steven Ka Shing Yau, Zijuan Lin, et al. 2024.
MetaGPT: Meta Programming for A Multi-Agent Collaborative Framework.
In \textit{Proceedings of the Twelfth International Conference on Learning Representations (ICLR '24)}.

\bibitem{hou2024llmranker}
Yupeng Hou, Junjie Zhang, Zihan Lin, Hongyu Lu, Ruobing Xie, Julian McAuley, and Wayne Xin Zhao. 2024.
Large Language Models are Zero-Shot Rankers for Recommender Systems.
In \textit{Proceedings of the 46th European Conference on Information Retrieval (ECIR '24)}. 364--381.

\bibitem{huang2023interecagent}
Xu Huang, Jianxun Lian, Yuxuan Lei, Jing Yao, Defu Lian, and Xing Xie. 2023.
Recommender AI Agent: Integrating Large Language Models for Interactive Recommendations.
\textit{arXiv preprint arXiv:2308.16505}.

\bibitem{kang2018sasrec}
Wang-Cheng Kang and Julian McAuley. 2018.
Self-Attentive Sequential Recommendation.
In \textit{Proceedings of the 2018 IEEE International Conference on Data Mining (ICDM '18)}. 197--206.

\bibitem{lewis2020rag}
Patrick Lewis, Ethan Perez, Aleksandra Piktus, Fabio Petroni, Vladimir Karpukhin, Naman Goyal, Heinrich K{\"u}ttler, Mike Lewis, Wen-tau Yih, Tim Rockt{\"a}schel, et al. 2020.
Retrieval-Augmented Generation for Knowledge-Intensive NLP Tasks.
In \textit{Advances in Neural Information Processing Systems (NeurIPS '20)}. 9459--9474.

\bibitem{li2024generative}
Lei Li, Yongfeng Zhang, Dugang Liu, and Li Chen. 2024.
Large Language Models for Generative Recommendation: A Survey and Visionary Discussions.
In \textit{Proceedings of the 2024 Joint International Conference on Computational Linguistics, Language Resources and Evaluation (LREC-COLING '24)}. 10146--10159.

\bibitem{li2024camel}
Guohao Li, Hasan Abed Al Kader Hammoud, Hani Itani, Dmitrii Khizbullin, and Bernard Ghanem. 2024.
CAMEL: Communicative Agents for ``Mind'' Exploration of Large Language Model Society.
In \textit{Advances in Neural Information Processing Systems (NeurIPS '24)}.

\bibitem{lin2024can}
Jianghao Lin, Xinyi Dai, Yunjia Xi, Weiwen Liu, Bo Chen, Hao Zhang, Yong Liu, Chuhan Wu, Xiangyang Li, Chenxu Zhu, Huifeng Guo, Yong Yu, Ruiming Tang, and Weinan Zhang. 2024.
How Can Recommender Systems Benefit from Large Language Models: A Survey.
\textit{ACM Transactions on Information Systems} (2024).

\bibitem{liu2024llmers}
Qidong Liu, Xiangyu Zhao, Yuhao Wang, Yejing Wang, Zijian Zhang, Yuqi Sun, Xiang Li, Maolin Wang, Pengyue Jia, Chong Chen, Wei Huang, and Feng Tian. 2024.
Large Language Model Enhanced Recommender Systems: A Survey.
\textit{arXiv preprint arXiv:2412.13432}.

\bibitem{lubos2024llmexplain}
Petr Lubos, Ladislav Peska, and Patrik Slav{\'\i}k. 2024.
User Evaluation of LLM-Generated Explanations for Recommender Systems.
In \textit{Proceedings of the 29th International Conference on Intelligent User Interfaces (IUI '24)}. 597--608.

\bibitem{ni2019amazon}
Jianmo Ni, Jiacheng Li, and Julian McAuley. 2019.
Justifying Recommendations using Distantly-Labeled Reviews and Fine-Grained Aspects.
In \textit{Proceedings of the 2019 Conference on Empirical Methods in Natural Language Processing (EMNLP '19)}. 188--197.

\bibitem{openai2023gpt4}
OpenAI. 2023.
GPT-4 Technical Report.
\textit{arXiv preprint arXiv:2303.08774}.

\bibitem{qiu2025gcrs}
Zhangchi Qiu, Zehui Wang, Jianan Wang, and Alan Wee-Chung Liew. 2025.
Graph Retrieval-Augmented LLM for Conversational Recommendation Systems.
\textit{arXiv preprint arXiv:2503.06430}.

\bibitem{reimers2019sentencebert}
Nils Reimers and Iryna Gurevych. 2019.
Sentence-BERT: Sentence Embeddings using Siamese BERT-Networks.
In \textit{Proceedings of the 2019 Conference on Empirical Methods in Natural Language Processing (EMNLP '19)}. 3982--3992.

\bibitem{rendle2009bpr}
Steffen Rendle, Christoph Freudenthaler, Zeno Gantner, and Lars Schmidt-Thieme. 2009.
BPR: Bayesian Personalized Ranking from Implicit Feedback.
In \textit{Proceedings of the Twenty-Fifth Conference on Uncertainty in Artificial Intelligence (UAI '09)}. 452--461.

\bibitem{said2024explaining}
Alan Said. 2024.
On Explaining Recommendations with Large Language Models: A Review.
\textit{Frontiers in Big Data} 7 (2024), 1505284.

\bibitem{shu2024kgllm}
Yutong Shu, Peng Zhang, Yifan Li, and Chuang Zhang. 2024.
Knowledge Graph-Enhanced LLM for Multi-hop Link Prediction.
\textit{arXiv preprint arXiv:2402.12345}.

\bibitem{silva2024chatgpt}
Itallo Silva, Leandro Marinho, Alan Said, and Martijn Willemsen. 2024.
Leveraging ChatGPT for Automated Human-Centered Explanations in Recommender Systems.
In \textit{Proceedings of the 29th International Conference on Intelligent User Interfaces (IUI '24)}. 597--608.

\bibitem{touvron2023llama2}
Hugo Touvron, Louis Martin, Kevin Stone, Peter Albert, Amjad Almahairi, Yasmine Babaei, Nikolay Bashlykov, Soumya Batra, Prajjwal Bhargava, Shruti Bhosale, et al. 2023.
Llama 2: Open Foundation and Fine-Tuned Chat Models.
\textit{arXiv preprint arXiv:2307.09288}.

\bibitem{grattafiori2024llama3}
Aaron Grattafiori, Abhimanyu Dubey, Abhinav Jauhri, Abhinav Pandey, Abhishek Kadian, Ahmad Al-Dahle, Aiesha Letman, Akhil Mathur, Alan Schelten, Alex Vaughan, et al. 2024.
The Llama 3 Herd of Models.
\textit{arXiv preprint arXiv:2407.21783}.

\bibitem{wang2019kgat}
Xiang Wang, Xiangnan He, Yixin Cao, Meng Liu, and Tat-Seng Chua. 2019.
KGAT: Knowledge Graph Attention Network for Recommendation.
In \textit{Proceedings of the 25th ACM SIGKDD International Conference on Knowledge Discovery \& Data Mining (KDD '19)}. 950--958.

\bibitem{wang2021kgin}
Xiang Wang, Tinglin Huang, Dingxian Wang, Yancheng Yuan, Zhenguang Liu, Xiangnan He, and Tat-Seng Chua. 2021.
Learning Intents behind Interactions with Knowledge Graph for Recommendation.
In \textit{Proceedings of the Web Conference 2021 (WWW '21)}. 878--887.

\bibitem{wang2023recmind}
Yancheng Wang, Ziyan Jiang, Zheng Chen, Fan Yang, Yingxue Zhou, Eunah Cho, Xing Fan, Xiaojiang Huang, Yanbin Lu, and Yingzhen Yang. 2023.
RecMind: Large Language Model Powered Agent for Recommendation.
\textit{arXiv preprint arXiv:2308.14296}.

\bibitem{wang2024macrec}
Zhefan Wang, Yuanqing Yu, Wendi Yu, Weizhi Ma, and Min Zhang. 2024.
MACRec: A Multi-Agent Collaboration Framework for Recommendation.
\textit{arXiv preprint arXiv:2402.15235}.

\bibitem{wang2025kragrec}
Shijie Wang, Hangyu Guo, Zhibo Cai, Yongwei Zhao, Yubin Bao, and Ge Yu. 2025.
Knowledge Graph Retrieval-Augmented Generation for LLM-based Recommendation.
In \textit{Proceedings of the 63rd Annual Meeting of the Association for Computational Linguistics (ACL '25)}.

\bibitem{wu2023autogen}
Qingyun Wu, Gagan Bansal, Jieyu Zhang, Yiran Wu, Beibin Li, Erkang Zhu, Li Jiang, Xiaoyun Zhang, Shaokun Zhang, Jiale Liu, et al. 2023.
AutoGen: Enabling Next-Gen LLM Applications via Multi-Agent Conversation.
\textit{arXiv preprint arXiv:2308.08155}.

\bibitem{wu2023survey}
Likang Wu, Zhi Zheng, Zhaopeng Qiu, Hao Wang, Hongchao Gu, Tingjia Shen, Chuan Qin, Chen Zhu, Hengshu Zhu, Qi Liu, Hui Xiong, and Enhong Chen. 2023.
A Survey on Large Language Models for Recommendation.
\textit{arXiv preprint arXiv:2305.19860}.

\bibitem{xie2024llmpkg}
Xie Liu, Chen Zhang, Xiangnan He, and Fuli Feng. 2024.
Enabling Explainable Recommendation in E-commerce with LLM-powered Product Knowledge Graph.
In \textit{IJCAI Workshop on Knowledge Graphs and LLMs}.

\bibitem{yao2023react}
Shunyu Yao, Jeffrey Zhao, Dian Yu, Nan Du, Izhak Shafran, Karthik Narasimhan, and Yuan Cao. 2023.
ReAct: Synergizing Reasoning and Acting in Language Models.
In \textit{Proceedings of the Eleventh International Conference on Learning Representations (ICLR '23)}.

\bibitem{yelp2023dataset}
Yelp. 2023.
Yelp Open Dataset.
\url{https://www.yelp.com/dataset}.

\bibitem{zhang2020explainable}
Yongfeng Zhang, Xu Chen, Qingyao Ai, Liu Yang, and W. Bruce Croft. 2020.
Towards Conversational Recommendation over Multi-Type Dialogs.
In \textit{Proceedings of the 58th Annual Meeting of the Association for Computational Linguistics (ACL '20)}. 1036--1049.

\bibitem{zhang2024agent4rec}
An Zhang, Yuxin Chen, Leheng Sheng, Xiang Wang, and Tat-Seng Chua. 2024.
On Generative Agents in Recommendation.
In \textit{Proceedings of the 47th International ACM SIGIR Conference on Research and Development in Information Retrieval (SIGIR '24)}. 1807--1817.

\bibitem{zhang2024agentcf}
Junjie Zhang, Yupeng Hou, Ruobing Xie, Wenqi Sun, Julian McAuley, Wayne Xin Zhao, Leyu Lin, and Ji-Rong Wen. 2024.
AgentCF: Collaborative Learning with Autonomous Language Agents for Recommender Systems.
In \textit{Proceedings of the ACM Web Conference 2024 (WWW '24)}. 3876--3887.

\bibitem{zhu2024agentrecsys}
Xi Zhu, Yu Wang, Hang Gao, Wujiang Xu, Chen Wang, Zhiwei Liu, Kun Wang, Mingyu Jin, Linsey Pang, Qingsong Wen, Philip Yu, and Yongfeng Zhang. 2024.
Recommender Systems Meet Large Language Model Agents: A Survey.
\textit{arXiv preprint arXiv:2411.00114}.

\bibitem{zhu2025kgrag}
Xiangrong Zhu, Yuexiang Xie, Yi Liu, Yaliang Li, and Wei Hu. 2025.
Knowledge Graph-Guided Retrieval Augmented Generation.
\textit{arXiv preprint arXiv:2502.06864}.

\end{thebibliography}

\appendix

\section{Prompt Templates}

We provide the key prompt templates used by MATRAG agents.

\subsection{User Modeling Agent Prompt}

\begin{verbatim}
You are a User Modeling Agent. Analyze 
the user's interaction history and 
extract structured preference signals.

User History:
{interaction_history}

Extract:
1. Explicit preferences (stated likes/dislikes)
2. Implicit preferences (inferred from behavior)
3. Contextual factors (time, device, session)
4. Preference evolution (temporal patterns)

Output as structured JSON.
\end{verbatim}

\subsection{Explanation Agent Prompt}

\begin{verbatim}
You are an Explanation Agent. Generate
a transparent, grounded explanation for
the recommendation.

Recommended Item: {item}
User Profile: {user_profile}
Reasoning Chain: {reasoning_chain}
Retrieved Knowledge: {kg_subgraph}

Generate an explanation that:
1. Cites specific evidence from knowledge
2. Connects to user preferences explicitly
3. Is honest about recommendation rationale
4. Uses natural, accessible language
\end{verbatim}

\section{Additional Experimental Results}

\subsection{Performance by User Activity Level}

Table~\ref{tab:activity} shows MATRAG's performance across users with different activity levels.

\begin{table}[h]
\caption{HR@10 by user activity level on Amazon Electronics.}
\label{tab:activity}
\centering
\small
\begin{tabular}{l|ccc}
\toprule
\textbf{Method} & Low & Medium & High \\
\midrule
LLMRank & 0.298 & 0.451 & 0.562 \\
MACRec & 0.341 & 0.478 & 0.589 \\
\textbf{MATRAG} & \textbf{0.412} & \textbf{0.534} & \textbf{0.628} \\
\bottomrule
\end{tabular}
\end{table}

MATRAG shows the largest improvements for low-activity users (+38.3\% over LLMRank), demonstrating that knowledge graph augmentation effectively addresses cold-start challenges.

\section{Reproducibility}

We will release code, data splits, and model checkpoints upon publication to ensure reproducibility.

\end{document}